\documentclass[12pt]{iopart} 
\usepackage{iopams}
\usepackage{cite}

\usepackage[utf8]{inputenc}
\usepackage{graphicx}
\usepackage{xcolor}

\usepackage{soul}  

\begin{document}

\title[Uniform quantized electron gas]{Uniform quantized electron gas}         
\author{Johan S. H\o ye$^1$ and  Enrique
  Lomba$^2$}

\address{$^1$Institutt for Fysikk,  NTNU, N-7491 Trondheim, Norway \\
  $^2$Instituto de Qu{\' \i}mica F{\' \i}sica Rocasolano, CSIC,
  Calle Serrano 119, E-28026 Madrid, Spain }

\date{\today}          
\begin{abstract}

In this work we study the correlation energy of the quantized electron
gas of uniform density at temperature $T=0$. To do so we utilize
methods from classical statistical mechanics. The basis for this is
the Feynman path integral for the partition function of quantized
systems. With this representation the quantum mechanical problem can
be interpreted as, and is equivalent to, a classical polymer problem in
four dimensions where the fourth dimension is imaginary time. Thus
methods, results, and properties obtained in the statistical mechanics
of classical fluids can be utilized. From this viewpoint we recover
the well known RPA (random phase approximation). Then to improve it we 
in this work modify the RPA by requiring the corresponding
correlation function to be such that electrons with equal spins can
not be on the same position. Numerical evaluations are compared with
well known results of a standard parameterization of Monte Carlo
correlation energies.

\end{abstract}
\maketitle

\submitto{\JPCM}

\bigskip
\section{Introduction} 
\label{sec1}      

It is a pleasure for us to contribute this article to a memorial issue
for Professor George Stell. We both have had an extensive scientific
collaboration with him in the research field of statistical mechanics
to which he was deeply devoted. For one of us (JSH) this cooperation
lasted for more than 30 years since 1973. As sketched below,
developments, insights, and results through those years form a basis
for the present work on the quantized electron gas.  

The initial common interest back in 1973 was in the $\gamma$-ordering
introduced by Hemmer \cite{hemmer64} and by Lebowitz, Stell, and Baer
\cite{lebowitz65}. The $\gamma$ is the inverse range of a perturbing
interaction added to a reference system, e.g. a fluid consisting of
hard spheres. The $\gamma$-ordering is a resummation of the well known
Mayer graph expansion in classical statistical mechanics. In this way
one immediately finds a leading correction to mean field theories,
like the van der Waals theory of fluids; and it is not restricted to
low density as the usual low density virial expansion is. However,
there are problems as divergences show up in the critical region of
the gas-liquid phase transition. 

With this background we initiated works on the statistical mechanics
of polar fluids with dipole-dipole pair interaction. This was extended
to the MSA (mean spherical approximation) of fluids where solutions of
the Ornstein-Zernike equation were studied \cite{lebowitz66}. The MSA was a modification
of the SM (spherical model) of spin
systems. \cite{berlin53} It turned out to correspond to the
leading order of 
$\gamma$-ordering beyond mean field with the additional constraint
that the corresponding pair correlation function should fulfill the
exact core condition for hard spheres where they cannot overlap. 

Further it was realized that a fluid consisting of polarizable
particles with fluctuating dipole moments modeled as harmonic
oscillators, could be quantized using the methods of classical
statistical mechanics \cite{hoye81}. In this case quantization was
restricted to the internal oscillatory motions of dipole moments, not
including particle positions. This extension to quantum systems was based on
Feynmann's path integral formalism \cite{feynman53}. The path integral
turns out to be equivalent to a classical polymer problem (or random
walk) problem in four dimensions, the fourth dimension being imaginary
time. These polymers are periodic in the fourth dimension of length
$\beta=1/(k_B T)$ where $k_B$ is Boltzmann's constant and $T$ is
temperature. This classical polymer interpretation can also be used
for systems consisting of fermions as well as bosons
\cite{chandler81,hoye94}. 

By study of and work on Casimir forces between closely separated
dielectric or metal plates, it was seen that these forces can be
interpreted as induced between quantized oscillating dipolar moments
of particles \cite{hoye98}. Forces between "metal" plates filled with classical
charged particles interacting with the Coulomb interaction were also considered \cite{hoyebrevik}. In this connection it was noted that the
quantized electromagnetic field, since it is a linear theory, can be
fully eliminated to be replaced by time-dependent (radiating) dipolar
interactions. With static dipolar interactions, valid for short
distances, the Casimir forces are the same as induced van der Waals
forces. In this respect the quantized polarizable fluid of Ref.~\cite{hoye81} was extended to include radiating dipolar interactions, and corrections to van der Waals forces or the free energy were obtained for the fluid in bulk \cite{waage13}.

Standard methods to study interacting many particle systems are the
Hartree-Fock theory or DFT (density functional theory)  of quantum
mechanics \cite{hohenberg64}. In view of the path integral formalism this can be mapped
again onto a classical
polymer problem  \cite{hoye10}. In the framework  of $\gamma$-ordering, free
(non-interacting) fermions (or bosons) constitute the reference system. (From
this viewpoint it is an ideal gas of polymers tied together into coils
of varying numbers of turns. For fermions an even number of turns give
negative number densities \cite{hoye94}.) In general the quantum
fluid, like electrons in molecules, has a nonuniform  density due to the
external potential set up by the atomic nuclei and the mean field
produced by the interacting electrons. This polymer problem in mean
field can be solved by means of the corresponding Schr\"{o}dinger
equation. This serves as a mean field solution to the statistical
mechanical polymer problem, an ideal polymer gas in an external mean
field. Since free fermions (or bosons) are correlated, there will also
be energy contributions due to the reference system correlations. This
is the well known exchange energy. 

However, the pair interactions induce additional effects  that
contribute to a correlation energy. These correlations have been difficult to
handle. They are non-local in nature, and varying methods have been
used to estimate the corresponding correlation energy \cite{fulde}. 
Broad overviews are given in the articles by Burke and Becke \cite{burke12}. With our statistical mechanical development we will recover the RPA for quantized many-body systems \cite{pines52,lein00}. Properties found from this development will be utilized in the effort to improve upon the RPA.

In Casimir theory, induced forces follow from the dielectric properties
of media. These properties again are related to molecular
polarizabilities. And by elimination of the electromagnetic field (in
thermal equilibrium) as mentioned above, these forces follow from
induced interactions between fluctuating dipole moments. But then this
energy will be part of molecular energies too and should be included
\cite{hoye10}. Thus the correlation energy can be identified with the
Casimir energy, or in the electrostatic case the van der Waals energy,
in bulk. (The standard Casimir problem, however, is more limited, as
only the induced free energy due to the interaction between the plates
is needed to obtain the force.) This again corresponds to a contribution to
leading order in $\gamma$-ordering. Here it can be noted that the
statistical mechanical graph structure is valid and thus also
applicable for non-uniform systems, like the system of two plates of
the Casimir problem. But in the  general situation the numerical problems will be much more
demanding. A recent review of van der Waals forces and use of vdW-DF method to account for van der Waals interactions in DFT is given by Berland et al.~\cite{berland15}.

Here we will only consider the uniform case, but properties and
results found may be applicable to the general situation. On the basic
microscopic level the electron gas is not a dielectric medium, but
consists of charged particles interacting via the Coulomb interaction
where we will limit ourselves to the electrostatic case. The theory
considered and established in the following sections will be followed
by numerical evaluation of explicit results which will be compared
with available results from computer simulations. 

In Section \ref{sec2} the leading order beyond mean field in
$\gamma$-ordering is considered at temperature $T=0$. It turns out
that this is equivalent to the well known RPA (random phase
approximation). Explicit RPA equations for the uniform electron gas
are established. As expected, when compared with simulations there are
clear deviations. 

In Section \ref{sec3}, a parameterized effective interaction, which modifies the Coulomb
interaction for small distances, is considered. This is
inspired by properties of the direct correlation function for
classical systems, which in the MSA follows the pair potential
outside the hard core diameter, but is dictated by the hard core
condition inside. For fermions, like electrons, particles with equal
spins cannot be at the same position. This gives a condition to
determine parameters that define the effective interaction. 

In Section \ref{sec4} the Fourier transform in space and imaginary
time for the correlation function of the free uniform electron gas is
given. Then the correlation functions where induced correlations are
present, are established.  

In Section \ref{sec5} the equations established are extended to those of
a two-component mixture of electrons with $\pm\frac{1}{2}$ spins. Such
an extension is necessary to be able to take into account that pairs
of equal and unequal spins will have effective interactions that must
be different for short distances. 

In Section \ref{sec6}, we will explore  various effective or cut Coulomb interactions 
 to be used in the numerical investigations of this work. 

In Section \ref{sec_num} we introduce our most significant
results and discuss them in the context of well known approximations
for the correlation energy of the quantized electron gas. Future
prospects and   conclusions are presented in Section \ref{sec_con}.

\section{Random phase approximation} 
\label{sec2} 
 
The Fourier transform (in space) of the pair correlation function or structure factor for free bosons and fermions is given by \cite{hoye94,hoye10A}
\begin{equation}
{\tilde S}(\lambda,k)=\frac{\zeta}{(2\pi)^3}\int\frac{{\tilde F}_\lambda (k'){\tilde F}_{\beta-\lambda}(k'')}{(1\pm \zeta X)(1\pm \zeta Y)}\,d{\bf k'}
\label{1}
\end{equation}
where ${\bf k''}={\bf k}-{\bf k'}$ and $\lambda=it/\hbar$ ($t$ is time), $0<\lambda<\beta$. The $\beta=1/(k_B T)$ where $k_B$ is Boltzmann's constant and $T$ is temperature. Further
\begin{equation}
F_\lambda(k)=\exp{(-\lambda E(k))},
\quad X=F_\beta (k'), \quad {\rm and} \quad Y=F_\beta (k'')
\label{2}
\end{equation}
where $E(k)$ is the energy of particles with mass $m$, i.~e.
\begin{equation}
E(k)=\frac{1}{2m}(\hbar k)^2.
\label{3}
\end{equation}
Finally $\zeta=e^{\beta\mu}$ where $\mu$ is the chemical potential. In Eq.~(\ref{1})
the plus sign is for fermions while the minus sign is for bosons. By Fourier transform in imaginary time $\hat S(K,k)=\int_0^\beta \tilde S(\lambda,k)e^{iK\lambda}\,d\lambda$ one obtains 
\begin{equation}
{\hat S}(K,k)=\frac{\zeta}{(2\pi)^3}\int\frac{1}{iK+\Delta}\frac{X-Y}{(1\pm \zeta X)(1\pm \zeta Y)}\,d{\bf k'}
\label{4}
\end{equation}
where $\Delta=E(k'')-E(k')$ and $K=2\pi n/\beta$ with $n$ integer are the Matsubara frequencies. 

Now the charged particles (electrons) interact via the Coulomb
interaction $\psi(r)=e^2/(4\pi \varepsilon_0 r)$ (in SI units) whose
Fourier transform is 
\begin{equation}
\tilde\psi(k)=\frac{e^2}{\varepsilon_0 k^2}.
\label{5}
\end{equation}
The $-e$ is the electron charge while $\varepsilon_0$ is the permitivity of vacuum.

The non-interacting particles form the reference system electron gas
of fermions. To obtain the contribution to the free energy from
the pair interaction we follow the $\gamma$-ordering scheme developed
for classical fluids \cite{hemmer64,lebowitz65}. The $\gamma$ is the
perturbing parameter, and it is the inverse range of attraction. (For
Coulomb interaction the $\gamma$ may be regarded as the inverse range
of the shielded Coulomb interaction that follows from summation of
chain bond graphs that also lead to the well known Debye-H\"{u}ckel
theory.) 

The leading contribution to Helmholtz free energy $\Delta F$ now follows from summation of the ring graphs with potential bonds and is given by \cite{hoye11}
\begin{equation}
-\beta \Delta F=-\frac{1}{2(2\pi)^3}\sum_K\int\,d{\bf k}\ln[1-g\hat S(K,k)(-\tilde\psi(k)]
\label{6}
\end{equation}
where $g=2$ takes into account degeneracy of up and down electron spins (that for free electrons are uncorrelated).

By expansion of the logarithm the linear term can be separated out as
\begin{equation}
-\beta \Delta F_{ex}=-\frac{g\beta}{2(2\pi)^3}\int\,d{\bf k}\tilde S(0,k)(-\tilde\psi(k)
\label{7}
\end{equation}
since $(1/\beta)\sum_K\hat S(K.k)=\tilde S(0,k)$ (i.e. $\lambda=0$). The $\Delta F_{ex}$ is the well known exchange energy. Remaining free energy is the non-local part, the correlation energy, due to induced correlations from the pair interactions. We are interested in the ground state energy at $T=0$ by which $\beta\rightarrow\infty$, and $(1/\beta)\sum_K\rightarrow(1/(2\pi))\int\,dK$. At $T=0$ entropy is zero for quantum systems by which internal energy is the same as Helmholtz free energy. Thus the correlation energy per unit volume becomes
\begin{equation}
-\beta \Delta F_c=-\frac{1}{2(2\pi)^3}\frac{1}{2\pi}\int\,dK\int\,d{\bf k}[\ln(1-\hat A(K))-\hat A(K)],
\label{8}
\end{equation}
\begin{equation}
\hat A(K)=-g\hat S(K,k)(-\tilde\psi(k)).
\label{8b}
\end{equation}
This expression for $\Delta F_c$ turns out to be the same as the RPA (random phase approximation) \cite{pines52,lein00}. 

With known $\hat S(K,k)$ Eq.~(\ref{8}) can be integrated
numerically. In Ref.~\cite{hoye11} a very crude approximation $\hat
S(K,k)=2\Delta\tilde S(0,k)/(K^2+\Delta^2)$ (with $\Delta\sim k$) was
used to enable analytic integration with respect to $K$ to perform
further analysis. However, integral (\ref{4}) for $\hat S(K,k)$ can be
evaluated analytically for $T=0$. The result is given by the Lindhard
function $\chi^0$, Eq.~(28) of Ref.~\cite{lein00}, which in our
notation below is 
\begin{equation}
g\hat S(K,k)=-\chi^0=\frac{mk_f}{2\pi^2\hbar^2}f(Q,x)
\label{9}
\end{equation}
\begin{eqnarray}
\nonumber
\displaystyle
f(Q,x)&=&-\left[\frac{Q^2-x^2-1}{4Q}\ln\left(\frac{x^2+(Q+1)^2}{x^2+(Q-1)^2}\right)\right.\\
&&\left.-1+x\arctan\left(\frac{1+Q}{x}\right)+x\arctan\left(\frac{1-Q}{x}\right)\right]
\label{10}
\end{eqnarray}
with
\begin{equation}
x=\frac{mK}{\hbar^2k k_f}=\frac{K}{4\mu Q},\quad \mu=\frac{(\hbar k_f)^2}{2m}, \quad Q=\frac{k}{2 k_f}
\label{11}
\end{equation}
where $\mu$ is the Fermi energy and $k_f$ is the Fermi wave vector.

With this and interaction (\ref{5}) the quantity $\hat A(K)$ in expression (\ref{8}) for the energy can be expressed as
\begin{equation}
\hat A(K)=D\frac{f(Q,x)}{Q^2}, \quad D=\frac{m k_f}{2\pi^2\hbar^2}\frac{e^2}{\varepsilon_0(2k_f)^2}=\frac{3}{32}\left(\frac{\hbar\omega_p}{\mu}\right)^2.
\label{12}
\end{equation}
The $\omega_p$ is the plasma frequency
\begin{equation}
(\hbar\omega_p)^2=\frac{\rho e^2}{m\varepsilon}
\label{13}
\end{equation}
where $\rho$ is the number density of particles given by ($T\rightarrow 0$) \cite{hoye11}
\begin{equation}
\rho=\frac{g}{(2\pi)^3}\int\frac{\zeta X}{1+\zeta X}\, d{\bf k}\rightarrow\frac{g}{(2\pi)^3}\int_{k<k_f} d{\bf k}=\frac{4\pi g}{3(2\pi)^3}k_f^3.
\label{14}
\end{equation}

For the quantized electron gas at $T=0$ there is only one independent parameter. A standard choice is the parameter $r_s$ given by
\begin{equation}
\frac{4\pi}{3}(r_s a_0)^3=\frac{1}{\rho}, \quad a_0=\frac{4\pi\varepsilon_0\hbar^2}{me^2}
\label{15}
\end{equation}
where $a_0$ is the Bohr radius. With this definition, $r_s$ is unitless. Inserting for $a_0$ and $\rho$ one finds
\begin{equation}
r_s=12.0584\cdot D\quad \mbox{or}\quad D=0.082293\cdot r_s
\label{16}
\end{equation}
with $D$ given by Eq.~(\ref{12}). For the correlation energy per particle Eq.~(\ref{8}) can be rewritten as
\begin{equation}
f_c=\frac{\Delta F_c}{\rho}=12\int_0^\infty\tilde f_c(k)Q^2\,dQ
\label{17}
\end{equation}
\begin{eqnarray}
f_c(k)&=&\frac{1}{2}\frac{1}{2\pi}\int[\ln(1+\hat A(K))-\hat
  A(K)]\,dK\nonumber\\
&= &\frac{1}{\pi}\mu Q\int[\ln(1+\hat A(K))-\hat A(K)]\,dx
\label{18}
\end{eqnarray}
with $\rho$ given by Eq.~(\ref{14}) and $x$, $\mu$, and $Q$ given by Eq.~(\ref{11}). Since $\mu\sim k_f^2$ and $r_s\sim\rho^{-1/3}\sim k_f^{-1}$ it follows that
\begin{equation}
\mu=\mu(r_s)=\frac{\mu(r_s=1)}{r_s^2}=\frac{50.1\,\rm{eV}}{r_s^2}
\label{19}
\end{equation}
with $\hbar=1.054\cdot10^{-34}$\,Js and electron mass $m=9.11\cdot 10^{-31}$\,kg inserted ($1\,eV=1.602\cdot 10^{-19}$\,J). The $k_f$ follows from Eqs.~(\ref{14}) and (\ref{15}) with $e=1.602\cdot 10^{-19}$\,As and $\varepsilon=8.854\cdot10^{-12}$\,As/(Vm) inserted. (One also finds $\hbar\omega_p=(47.1\,\rm{eV}/r_s^{3/2})$.   

In Fig.~2 of Ref.~\cite{lein00} results for the distribution of energies ($Q=k/(2k_f)$)                 
\begin{equation}
\varepsilon_c(Q)=12 f_c(k)Q^2
\label{20}
\end{equation}
are plotted as function as function of $Q$ for $r_s=4$. There one of the curves is the one of the RPA which is reproduced by the expressions established above. These expressions will be a basis for the modifications performed below to improve results.

\section{Effective interaction} 
\label{sec3} 

For classical fluids various properties or conditions can be utilized
to  improve results. One such condition is the hard sphere condition
which we will focus upon here. This implies that particles can not
overlap within their hard core radii. This is expressed through the
pair correlation function $h(r)$ that can be required to have its
exact value -1 (with standard definition of $h(r)$) within the hard
core diameter. But the main problem is that the hard core influences
the correlation function in a highly non-trivial way outside the hard
core. The latter problem, however, is resolved by considering the
direct correlation function $c(r)$ where simple approximations can be
made \cite{ornstein14}. The $h(r)$ and $c(r)$ for classical fluids are related to each
other via the Ornstein-Zernike integral equation which is similar to
the Dyson type equation (\ref{24}) below (with its $\hat S(K,k)$
replaced with $\rho$). 

It is known that to a leading approximation the
$c(r)=-\beta\psi(r)$ where $\psi(r)$ is the perturbing
interaction. Further this becomes exact for large
$r\rightarrow\infty$. For small $r$ there will be deviations. In the
MSA (mean spherical approximation) these deviations are defined to be
inside the range of the hard core diameter. With this and the hard
core condition on $h(r)$ as boundary conditions the resulting $h(r)$
can be determined for all $r$. With known $h(r)$ the equation of state
can be obtained both via the internal energy route and the
compressibility route. Since MSA is not exact these routes give
different results, a measure of resulting inaccuracy. 

Then the SCOZA
(self-consistent Ornstein-Zernike approximation) was introduced where
the MSA was used with $\beta$ replaced by an effective inverse
temperature $\beta_e$ requiring both routes to thermodynamics to give
the same result\cite{scoza}. This resulted in a non-linear partial
differential equation. By numerical solution of this equation very
accurate results came out, also in the "non-classical" critical
region\cite{wilson74,scoza2}. A related fluid theory, the HRT
(hierarchical reference theory), based upon the renormalization group approach also
gives very accurate results\cite{hrt}. The critical indices of the
latter have by analytic and numerical work been found to be simple
numbers\cite{lomba14}. It is not ruled out that these indices are the
exact ones for fluids, lattice gases, and spin systems in three
dimensions. Anyway, it is thus demonstrated that the exact $c(r)$
deviates from its MSA form by some function the form of which is not
crucial for accurate results. 

Further the MSA was extended to polymers
where the reference system was hard spheres tied together into polymer
chains \cite{hoye04}. Again the MSA type $c(r)$ gave reasonably good
results. These properties are expected to carry over to the "polymer" path 
integral of quantized systems. Then in the RPA the $\psi(r)$ plays the role of the
$c(r)$. Again this should be exact as $r\rightarrow\infty$ while for
small $r$ there will be modifications that will replace $\psi(r)$ 

The latter polymer problem in 3 dimensions is similar to the polymer or random walk like
path integral of quantum mechanics in 4 dimensions. In the RPA the interaction
$\psi(r)$ plays the role of the $c(r)$. Again this should be exact as
$r\rightarrow\infty$ while for small $r$ there will be
modifications. Thus similar to classical fluids this will replace the
$\psi(r)$ with an effective interaction. Similar efforts have been done
earlier \cite{lein00}. A key problem in this context has been how to
specify the effective interaction more precisely.  

Thanks to the above connection to classical fluids, we can impose similar
conditions to the path integral "polymers" of quantized electrons
that interact. Electrons are fermions that behave as hard spheres in
the sense that they are not allowed to occupy the same site if 
their spins are equal. Thus one can apply a hard sphere condition on
the equal time pair correlation function for electrons with equal spins at the
same position. For slightly different positions the situation is less
clear as electrons start to overlap. However, one will expect that the
effective interaction must transform to the $\psi(r)$ in a smooth way
during a distance corresponding to the one between particles or the
inverse of the Fermi wave vector $k_f$ as $r$ increases. Due to the
repulsive Coulomb interaction one might expect the same hard core
condition for unequal spins. We will find, however, that the latter
fails especially when the parameter $r_s$ decreases. The reason is
apparently that quantum mechanically the wave function does not go
down to zero at $r=0$ with repulsive Coulomb interaction between
unequal spins. Due to increasing fermion pressure with density, this
overlap will increase with decreasing $r_s$. 

\section{Correlation function and core condition} 
\label{sec4}

The structure function of the free Fermi gas is given by Eq.~(\ref{4}) together with explicit expressions (\ref{9}) and (\ref{11}). The equal time pair correlation function $h_{ii}(r)$ ($i=1,2$ for $\pm\frac{1}{2}$ spins) is related to it as (at $T=0$)
\begin{equation}
S(0,r)=\rho_i\delta({\bf r})+ \rho_i^2 h_{ii}(r), \quad \rho_i=\frac{\rho}{2}.
\label{21}
\end{equation}
Explicitly one has
\begin{equation}
h_{ii}(r)=-(g(r))^2, \quad g(r)=\frac{3(\sin\gamma-\gamma\cos\gamma)}{\gamma^3}, \quad \gamma=k_f r.
\label{22}
\end{equation}
As $g(r)\rightarrow 1$ when $r\rightarrow 0$, the fermion condition $h_{ii}(0)=-1$ i fulfilled. For unequal uncorrelated spins $h_{12}(r)=0$. Further for the equal time Fourier transform ($g=2$)
\begin{equation}
g\tilde S(0,k)=
\left\{\begin{array}{cc}
\rho(\frac{3}{2}Q-\frac{1}{2}Q^3), &\quad Q<1\\
\rho ,                          & \quad  Q>1
\end{array} \right.
\label{23}
\end{equation}
with $Q$ and $\rho$ given by Eqs.~(\ref{11}) and(\ref{14}).

By $\gamma$-ordering the leading correction to the reference system pair correlation function are the graphs that form the chain bond expression. This also coincides with the RPA and is given by \cite {lein00}
\begin{equation}
g\hat\Gamma(K,k)=\frac{g\hat S(K,k)}{1+\hat A(K)}
\label{24}
\end{equation}
with $\hat A(K)$ given by Eq.~(\ref{12}). Here, so far, the  electron gas is considered as a one-component fluid by which an average of correlations between equal and unequal spins is obtained.

Now we can introduce the effective  interaction where the Coulomb
interaction is modified for small $r$. We will also keep it finite for
small $r\rightarrow 0$ although this may not be required since
singular behavior will be smoothed out with Eq.~(\ref{24}). In $k$-space
the modification is equivalent to a smooth cut for large $k$. Thus the
interaction (\ref{5}) is replaced by an effective interaction 
\begin{equation}
\tilde\psi_e(k)=\tilde\psi(k)L(Q), \quad L(0)=1.
\label{25}
\end{equation}
In $r$-space, where it turned out more convenient to find a suitable functional form, this can be written
\begin{equation}
\psi_e(r)=\psi(r)f(r), \quad f(r)\rightarrow 1\quad\rm{as}\quad r\rightarrow\infty.
\label{26}
\end{equation}
The $f(r)$ will contain a free parameter that may be determined by a core condition at $r=0$.

With the crude analytic evaluations made in Ref.~\cite{hoye11} it was noted that the "exact" energy distribution could be reproduced rather well by a suitable choice of $L(Q)$ after first modifying the crude approximation for $\hat S(K,k)$ (as given in text above Eq.~(\ref{9})) to fit the RPA of Fig.~2 in Ref.~\cite{lein00}.

Here we want to use the core condition to see to which extent "exact"
results from computer simulations can be recovered when implementing
it. With Eq.~(\ref{24}) one can impose an average hard core
condition for equal and unequal spins. From Eqs.~(\ref{21}) and
(\ref{22}) follows as $r\rightarrow 0$ 
\begin{equation}
gS(0,r)=\rho\delta({\bf r})-\frac{1}{2}\rho^2.
\label{27}
\end{equation}
This is 1/2 of a hard core condition for both equal and unequal spins
taken together. So a corresponding full hard core condition on the
average requires ($g=2$) 
\begin{equation}
\Gamma(0,0)-S(0,0)=-\frac{1}{4}\rho^2
\label{28}
\end{equation}

 The resulting energy distribution (\ref{20}) has been evaluated
 numerically using various forms of the function         $f(r)$ that
 cuts the potential. In order to evaluate (\ref{28}) one considers 
\begin{equation}
\Delta\hat\Gamma(K,k)=\hat\Gamma(K,k)-\hat S(K,k),
\label{29}
\end{equation}
and integrate to find
\begin{equation}
\Delta\tilde\Gamma(0,k)=\frac{1}{2\pi}\int\Delta\hat\Gamma(K,k)\,dK=\frac{4\mu Q}{2\pi}\int_{-\infty}^\infty\Delta\hat\Gamma(K,k)\,dx,
\label{30}
\end{equation}
with use of Eqs.~(\ref{9}) and (\ref{11}). Further
\begin{equation}
\Delta\Gamma(0,0)=\frac{1}{(2\pi)^3}\int\Delta\tilde\Gamma(0,k)\,d{\bf k}=24\left(\frac{\rho}{2}\right)\int_0^\infty\Delta\tilde\Gamma(0,k)Q^2\,dQ,
\label{31}
\end{equation} 
where also Eq.~(\ref{14}) is used. This together with condition (\ref{28}) will determine a free parameter in the effective potential (\ref{26}).

We have tried some different forms for the cut function $f(r)$ and
considered the case $r_s=4$ where exact results are available in
Ref.~\cite{lein00} in the form of the Perdew-Wang
parameterization\cite{PW} of Monte Carlo results. By use of the average core condition (\ref{28})
the results obtained are seen in Fig.~\ref{figexcq} for various effective cut interactions (see Section \ref{sec6} below). As seen from the
figure results obtained were near the exact curve, considerably
improving upon the RPA curve, but with some deviations especially for large $Q$. 

However, in this calculation we have not taken into account that equal
and unequal pairs of spins correlate differently. This means that effective
interactions have to be different to fulfill the core conditions in
more detail, especially the one for equal spins that is strict. Moreover,
it turned out, as mentioned earlier, that the strict core condition could not be maintained
for all electron densities and $q$-vales 
on unequal pairs of spins. This condition had to be abandoned as computations
finally would no longer converge for increasing density.   Study
of the Schr\"{o}dinger equation shows  that the repulsive Coulomb
potential can not prevent overlap of unequal particles. Due to
increasing fermion pressure this overlap increases for higher
density. 

\section{Mixture of electrons with ${\bf\pm1/2}$ spins } 
\label{sec5}

With a mixture of $\pm\frac{1}{2}$ spins the correlation function will be a $2\times 2$ matrix. To simplify we introduce the matrices
\begin{eqnarray}
\nonumber
{\bf M}_+&=&\frac{1}{2}\left(\begin{array} {cc}
1 & \;1\\
1&  \;1
\end{array} \right), \quad
{\bf M}_-=\frac{1}{2}\left(\begin{array} {cc}
\;1 & -1\\
-1&  \;\,1
\end{array} \right),\\
{\bf M}_+^2&=&{\bf M}_+, \quad {\bf M}_-^2={\bf M}_-, \quad {\bf M}_+ {\bf M}_-=0.
\label{40}
\end{eqnarray}
The reference system correlation function can now be written as
\begin{equation}
\hat S(K,k)\rightarrow \hat S(K,k)({\bf M}_++{\bf M}_-).
\label{41}
\end{equation}
Likewise the Coulomb interaction can be written as
\begin{equation}
\tilde\psi(k)\rightarrow 2\tilde\psi(k) {\bf M}_+.
\label{42}
\end{equation}
However, the effective or cut interaction will be different for equal and unequal pairs of spins
\begin{equation}
\tilde\psi_e(k)\rightarrow 2(\tilde\psi_+(k) {\bf M}_++\tilde\psi_-(k) {\bf M}_-)
\label{43}
\end{equation}
Finally the resulting correlation function becomes
\begin{equation}
\hat \Gamma(K,k)\rightarrow\hat \Gamma_+(K,k) {\bf M}_+ +\hat \Gamma_-(K,k) {\bf M}_-
\label{44}
\end{equation}
with
\begin{equation}
\hat \Gamma_\pm(K,k)=\frac{\hat S(K,k)}{1+\hat A_\pm(K)},
\label{45}
\end{equation}
\begin{equation}
\hat A_\pm(K)=-g\hat S(K,k)(-\tilde \psi_\pm (k))=D\frac{f(Q,x)}{Q^2}L_\pm(Q),\quad L_\pm(Q)=\frac{\tilde\psi_\pm(k)}{\tilde\psi(k)}.
\label{46}
\end{equation}
The latter expression is a generalization of Eqs.~(\ref{8}), (\ref{12}), and (\ref{25}) when properties (\ref{40}) for the matrix multiplications are used. With this the correlation energy per particle (\ref{18}) is modified to
\begin{eqnarray}
\nonumber
f_c(k)&=\frac{1}{\pi}\mu Q\left\{\int[\ln(1+\hat A_+(K))-\hat A_+(K)]\,dx \right.\\
&\left.+\int[\ln(1+\hat A_-(K))-\hat A_-(K)]\,dx\right\}.
\label{47}
\end{eqnarray}
This is used in the energy distribution $\varepsilon_c(Q)$ given by Eq.~(\ref{20}) as before.
Expression (\ref{47}) for the free energy assumes that it keeps its MSA form also with effective interaction as is the case with classical hard spheres. However, from some of the various approximations considered in Ref.~\cite{lein00}, we realize that this may be modified. But in the present investigation this will not be considered.

Like Eqs.~(\ref{29}) - (\ref{31}) one now has
\begin{eqnarray}
\nonumber
\Delta\hat\Gamma_\pm(K,k)&=&\hat\Gamma_\pm(K,k)-\hat S(K,k)\\
\label{48}
\Delta\tilde\Gamma_\pm(0,k)&=&\frac{4\mu Q}{2\pi}\int_{-\infty}^\infty\Delta\hat\Gamma_\pm(K,k)\,dx\\
\nonumber
\Delta\Gamma_\pm(0,0)&=&24\left(\frac{\rho}{2}\right)\int_0^\infty\Delta\tilde\Gamma_\pm(0,k)Q^2\,dQ
\end{eqnarray}

With expressions (\ref{9}), (\ref{11}), and (\ref{14}) for $g\hat S(K,k)$, $\mu$, and $\rho$ respectively Eq.~(\ref{45}) gives ($g=2$)
\begin{equation}
\frac{4\mu Q}{2\pi}\Delta\hat\Gamma_\pm(K,k)=\frac{3Q}{2\pi}\left(\frac{\rho}{2}\right)\frac{f(Q,x)}{1+A_\pm(K)}
\label{48a}
\end{equation}
by which
\begin{equation}
\Delta\Gamma_\pm(0,0)=-\frac{36}{\pi}\left(\frac{\rho}{2}\right)^2\int_0^\infty\int_{-\infty}^\infty\frac{f(Q,x)A_\pm(K)}{1+A_\pm(K)}\,dx \,Q^3\,dQ.
\label{48b}
\end{equation}
With a strict core condition also on unequal spins the core condition
like Eq.~(\ref{28}) would be modified into  
\begin{equation}
\Delta\Gamma_\pm(0,0)=\mp\frac{1}{4}\rho^2.
\label{49}
\end{equation}
With this Eq.~(\ref{44}) for the mixture correlation function together with Eq.~(\ref{27}) for $S(0,0)$ would give ($r\rightarrow 0$, $\rho_i=\rho/2$)
\begin{equation}
\Gamma_+(0,0){\bf M}_+ +\Gamma_-(0,0){\bf M}_-=\frac{\rho}{2}\delta({\bf r})-\left(\frac{\rho}{2}\right)^2\left(
\begin{array} {cc}
1 & \;1\\
1&  \;1
\end{array} \right)
\label{50}
\end{equation}
However, it turned out that in a quantum mechanical framework the
strict hard core condition on unequal spins was not valid as mentioned
at the end of the previous section, and the solution of (\ref{49}) can
not be found for high density. This implies that Eq. (\ref{49}) has to be
abandoned to be replaced by ($a<\rho^2/4$)
\begin{equation}
\Delta\Gamma_\pm(0,0)=\mp a \quad \mbox{or} \quad \Delta\Gamma_+(0,0)+\Delta\Gamma_-(0,0)=0
\label{51}
\end{equation}

\section{Cut interaction } 
\label{sec6}

As mentioned before, the interaction is to be cut in a smooth way and will contain one or
more parameters to be determined in terms of the core condition. In this work we have the exact
condition (\ref{51}), and two free parameters are needed. This
means that  some
freedom remains, in addition to the form of the functions used. In
particular, the functions considered in this investigation were of the form 
\begin{equation}
\psi_+(r)=\psi(r)f(r), \quad \psi(r)=\frac{1}{4\pi\varepsilon_0 r}
\label{60}
\end{equation}
with $f(r)$ equal to a sharp cutoff, a simple exponential, a smoothed
exponential (with zero slope at the origin), an error function and a
smooth Gaussian function (whose value and derivative vanish as
$r\rightarrow 0$), i.e.
\begin{equation}
H(x-1),\quad 1-e^{-x},\quad 1-(1+x/2)e^{-x},\quad {\rm erf}(x),\quad
{\rm and}\quad 1-(1+x^2)e^{-x^2}
\label{61}
\end{equation}
respectively. Here $x=2 k_f \kappa r$, where $\kappa$ is a free
parameter ($H(x-1)$ is a Heaviside function, and
erf$(x)=(2/\sqrt{\pi})\int_0^x \exp({-u^2})\,du$). The corresponding
Fourier transforms are given by $\psi_+(k)=(1/(\varepsilon_0k^2))L(Q)$ with $L(Q)$ equal to 
\begin{eqnarray}
  \label{cut}
L_{sharp}(Q)&=& \cos(Q/\kappa)\label{sharp}\\
L_{se}(Q)&=&\frac{\kappa^2}{Q^2+\kappa^2},\label{se}\\
L_{exp}(Q)&=&\left(\frac{\kappa^2}{Q^2+\kappa^2}\right)^2,\label{exp}\\
L_{erf}(Q)&=&\exp{(-Q^2/(2\kappa)^2)},\label{erf}\\
L_{gauss}(Q)&=&1-Q^2\left(\left[\frac{3}{2}-\frac{Q^2}{4\kappa^2}\right]\frac{D_+(Q/(2\kappa))}{\kappa Q}+\frac{1}{4\kappa^2}\right),
\label{62}
\end{eqnarray}
respectively, with $D_+(x)$ being Dawson's integral\cite{abramowitz}, i.e.
\begin{equation}
D_+(x)=\frac{1}{2}\int_0^\infty e^{-u^2/4}\sin(xu)\,du=e^{-x^2}\int_0^x e^{t^2}\,dt.
\label{63}
\end{equation}
The last equality follows as both expressions solve the differential
equation $y'+2xy=1$. A graphical representation of the various cut
interactions in Fourier space can be seen in Figure \ref{phicut}. 

We also have to specify $\psi_-(r)$, which is a function of short
range. With $\psi_-(r)=\psi(r) L_-(Q)$ the choice 
\begin{equation}
L_-(Q)=-\tau Q^2 (L(Q)))^2
\label{63b}
\end{equation}
with one additional free parameter $\tau$ was made. Effectively this
means that the previous Coulomb interaction of $\psi_+(r)$ is replaced
with a shielded one ($1/Q^2\rightarrow L(Q))$. This is of course a
crude approximation, but  $\psi_-(r)$ is less significant than
$\psi_+(r)$,  since the former only deals with the difference
between particle pairs of either equal or unequal spins. 

\section{Numerical results} 
\label{sec_num}

At the end of Sec.~\ref{sec4}, Fig.~\ref{figexcq} was found by using the average hard core condition (\ref{28}). Further it was remarked that after further considerations the strict hard core condition would not 
be valid for unequal spins. Likewise the corresponding strict condition (\ref{49}) should be replaced by condition (\ref{51}) with a free parameter $a<(\rho/2)^2$. So for further computations in this work we chose to use
\begin{equation}
\Delta\Gamma_\pm(0,0)=\mp\frac{1}{2}\left(\frac{\rho}{2}\right)^2.
\label{70}
\end{equation}
With this and expression (\ref{46}) for $A_\pm(K)$ the core condition for $\Delta\Gamma_+(0,0)$ can
be cast into ($L_+(Q,\kappa)=L(Q)$, $L(Q,\kappa,\tau)=L_-(Q)$) 
\begin{equation}
\frac{1}{2} - \frac{36}{\pi} \int\int \frac{D f^2(Q,x)L_+(Q,\kappa)}{Q^2 +D
  f(Q,x)L_+(Q,\kappa)}dx\, Q^3\,dQ = 0.
\label{71}
\end{equation}
with $f(Q,x)$ given by Eq.~(\ref{10}) and $L_+(Q)$ defined by
Eqs.~(\ref{cut})-(\ref{62}). This equation is solved for $\kappa$
using a simple Newton-Raphson scheme. Now, with $\kappa$ known, using
(\ref{63b}) for $L_-(Q,\tau,\kappa)$, the corresponding core condition
for $\Delta\Gamma_-(0,0)$ can be solved for $\tau$
\begin{equation}
\frac{1}{2} + \frac{36}{\pi} \int\int \frac{D f^2(Q,x)L_-(Q,\kappa,\tau)}{Q^2 +D
  f(Q,x)L_-(Q,\kappa, \tau)}dx\, Q^3\,dQ = 0.
\label{72}
\end{equation}
Now, with Eqs.~(\ref{20}) and (\ref{47}) the correlation energy contribution  can be calculated as 
\begin{eqnarray}
\varepsilon_c(Q) & = & \frac{12\mu}{\pi} Q^3\left[\int \left(\ln (1+
\hat{A}_+(Q,x)) -\hat{A}_+(Q,x)\right) dx\right.\nonumber\\
& & \left.+\int \left(\ln (1+ \hat{A}_-(Q,x)) -\hat{A}_-(Q,x)\right) dx\right]
\end{eqnarray}

The above integrals have been evaluated 
up to $k=16k_f$ (i.e.~$Q=8$) with 2000 points and $x=10$ with 1000 points. For $r_s=4$
 solution of the core conditions yields $\kappa = 1.241$ and $\tau
= 0.617$. With this we have obtained the results shown in Fig.~\ref{epsnew}. 
Here one first can note the good accuracy with which the Gaussian cut
reproduces the "exact" correlation energy distribution. 
However, a second feature to be pointed out in this figure is the
breakdown of the results of our approximation when the effective cut
potential used corresponds to the simple exponential cut, (\ref{se}), the
smooth exponential, (\ref{exp}), and the error function, (\ref{erf}),
as $k$ decreases. This breakdown
occurs in the $\Delta\Gamma_-(0,0)$ part of the core condition (\ref{51}).
By a closer study of numerical results we found the breakdown connected to the 
denominator of Eq.~(\ref{72}) which became zero  (as $L_-(Q)<0$). Thus the chosen core condition (\ref{70}) could no longer be fulfilled for these functions. By making the core condition parameter $a$ of Eq.~(\ref{51}) smaller the breakdown may be prevented. But the main problem here may be connected to the crude guess for the function form (\ref{63b}) chosen. Anyway,we did not pursue this problem
further in this work. Instead we investigated more closely the two other cut potentials.

Interestingly, when the cut
interactions vanish as $r\rightarrow 0$ --as is the case of the
sharp cut, (\ref{sharp}), and the Gaussian cut, (\ref{62})-- the core
condition can be solved. One observes that the
Gaussian cut reproduces accurately the ``exact'' results obtained from
the Perdew-Wang parameterization of the correlation
energy\cite{lein00,PW}, with small (around one percent) deviations
close to the minimum at $k = 0.57 k_f$. In what follows, we
will restrict our discussion to the results obtained with the Gaussian
cut effective interaction.

We can now proceed to calculate the full correlation energy as
\begin{equation}
E_{c}(r_s) = \int \varepsilon_c(Q) d Q
\end{equation}
for a series of $r_s$. These results are plotted in Figures \ref{fxc}
and \ref{difxc}. In Figure \ref{fxc} we first observe that no results
could be obtained for $r_s < 1$ when both core conditions are
solved, since $\kappa$ grows rapidly as $r_s$ shrinks and this again leads to
a breakdown of the core condition for $\Delta\Gamma_-(0,0)$. 
This is again similar to the breakdown already met for $r_s=4$
for the other cut interactions. This is consistent with the picture that by increasing
pressure (or density) unlike spins can overlap more and more by which the core condition parameter $a$ of Eq.~(\ref{51}) must decrease in some way. But again, we did not consider this problem further.
So instead we decided first to bypass this
problem fixing the $r_s=4$ value of $\kappa$ for $r_s < 4$, and retaining
core condition (\ref{51}) with
$\kappa$ constant and the parameter $a$ eliminated. This improves somewhat the results and
above all it can be solved all the way down to $r_s\rightarrow
0$. Interestingly, if we keep both $\kappa$ and $\tau$ fixed to their $r_s=4$ values for $r_s
< 4$, then the results agree remarkably well with the PW
parameterization. This is an important result, given the simplicity of
the approach presented here, that would only require the solution of
the core conditions for moderate electron densities. 

Now, we can
compare our results with those of other approximations in Figure
\ref{difxc}. The approximations included in the figure are the
adiabatic local density approximation\cite{lein00} (ALDA), the RPA, the
parameterization of Corradini et al.\cite{corradini}, the
approximation of Petersilka, Grossman and Gross\cite{pgg}(PGG) and the
local approximation of Richards and Ashcroft\cite{RA} (LRA). The full
approximation of Richards and Ashcroft\cite{RA} (RA) practically
coincides with the PW parameterization and it is not included. With
the exception of the full RA approximation, we observe that our
approximation with the two parameters fixed for $r_s
< 4$  clearly outperforms all other approximations, exhibiting
appreciable differences only at very high electron densities. 

\section{Conclusion} 
\label{sec_con}

We have studied the correlation energy of the quantized electron gas
at uniform density. In view of the path integral the well known RPA
was recovered. The RPA can be interpreted in terms of a classical
statistical mechanical polymer problem in four dimensions. Methods and
reasoning of classical systems are thus utilized by which the Coulomb
interaction is modified for short distances to be replaced by an
effective one. Then, as imposed via the pair correlation function,
particle pairs of equal spins are not allowed to be at the same
position. Numerical results that agree very well with computer
simulations, are found. However, this required some extra adjustments of coefficients as described in the previous section. So in view of this and the remark below Eq.~(\ref{47})
further investigations are needed to possibly better straighten out problems that appeared or make other improvements.
We expect that results obtained for the
effective interaction for various electron densities may be extended
to and utilized for the non-uniform electron gas in molecules and on
periodic lattices to possibly obtain more accurate correlation
energies.

\ack EL acknowledges financial support from the
Direcci\'on General de Investigaci\'on Cient\'{\i}fica y T\'ecnica
under Grant No. FIS2013-47350-C5-4-R.

\newpage

\begin{figure}[h!]
\centering
\includegraphics*[width=18cm, angle=0,clip]{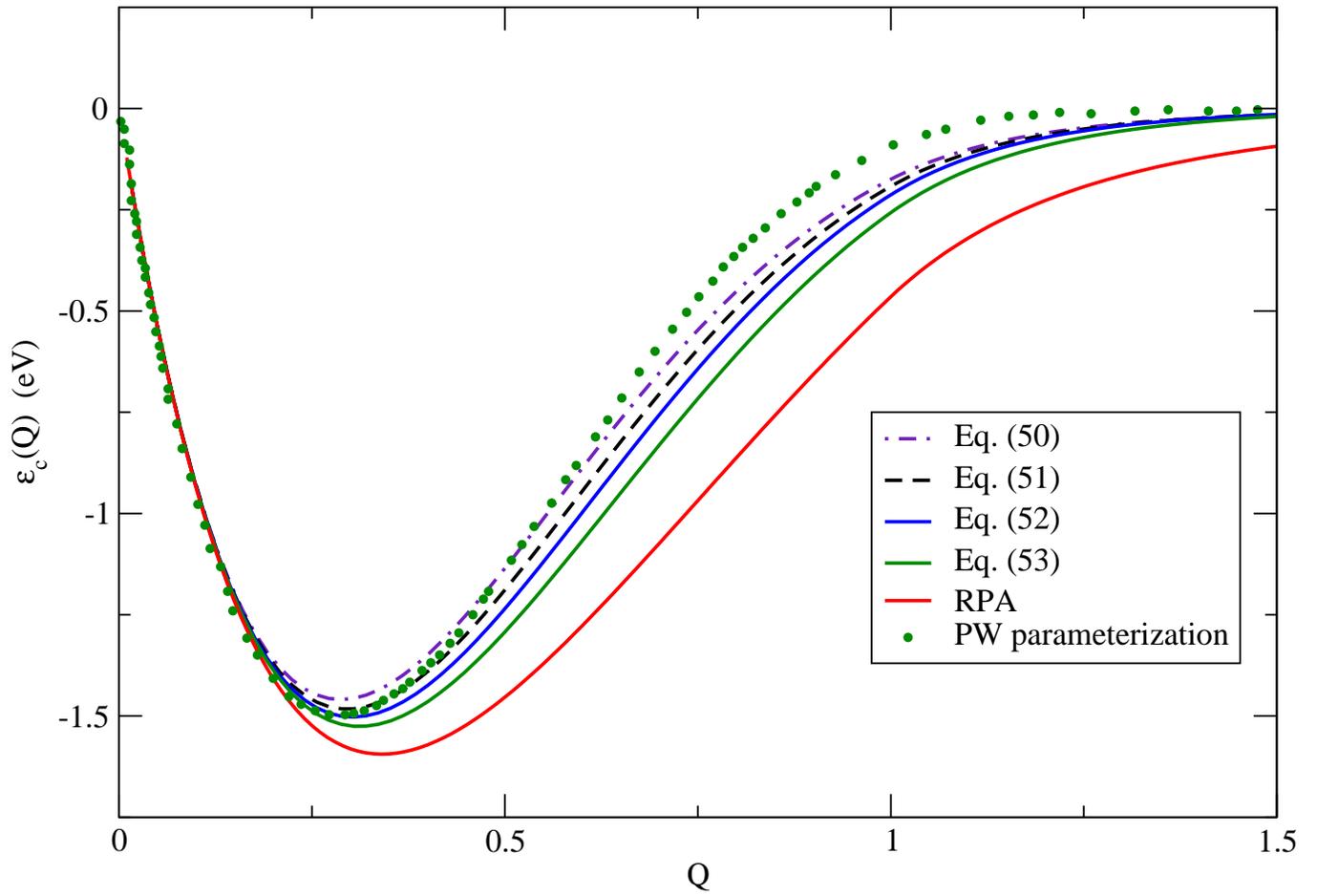}
\caption{Wave vector analysis of the correlation energy per electron
  calculated for the RPA and various cut interactions that follow from Eqs.~(\ref{se}) - (\ref{62}).
This is compared with the
  Perdew-Wang parameterization\cite{lein00,PW} computed for $r_s=4$
  . Note that $Q=k/(2k_f)$.} 
\label{figexcq}
\centering
\end{figure}

\newpage

\begin{figure}[h!]
\centering
\includegraphics*[width=18cm, angle=0,clip]{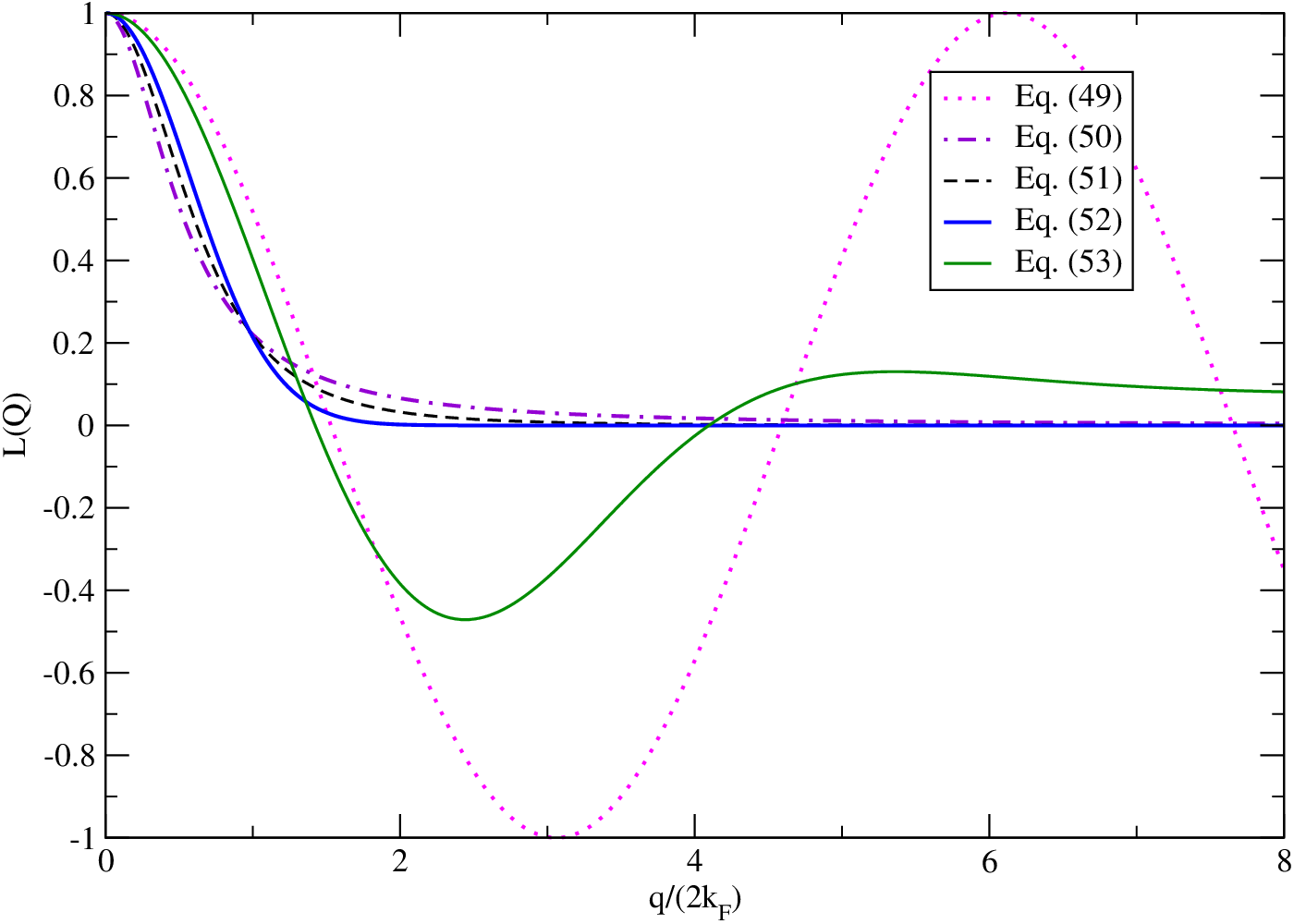}
\caption{Fourier transform of the cut effective
  potential relative to the one of the potential itself, $L(Q)$, calculated for the various functional forms used
  in this work, and listed in Eqs.~(\ref{cut})-(\ref{62}). Note that $Q=k/(2k_f)$.} 
\label{phicut}
\centering
\end{figure}

\begin{figure}[h!]
\centering
\includegraphics*[width=18cm, angle=0,clip]{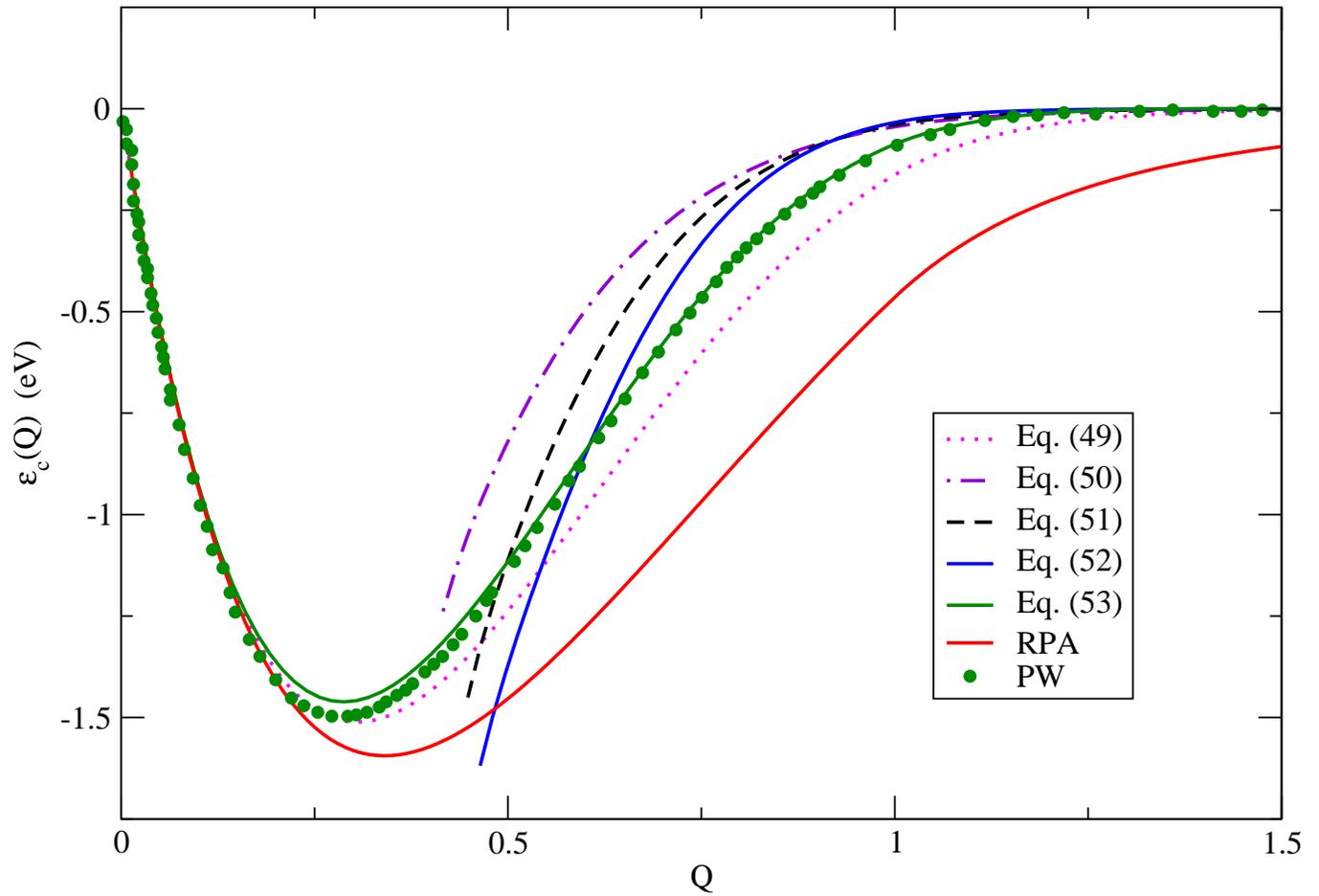}
\caption{Wave vector analysis of the correlation energy per electron,
  calculated for the RPA, using various effective cut interactions as
  defined in Eqs.(\ref{cut})-(\ref{62}), with the core conditions 
  implemented through Eqs.~(\ref{51}) and (\ref{70}) and using
  Eq.~(\ref{63b}). The ``exact'' results are computed from the
  Perdew-Wang parameterization\cite{lein00,PW}. Note for $Q\lesssim 0.45$ 
  computations fail for 3 of the curves. These calculations
  correspond to $r_s=4$. Note that $Q=k/(2k_f)$.} 
\label{epsnew}
\centering
\end{figure}

\begin{figure}[h!]
\centering
\includegraphics*[width=18cm, angle=0,clip]{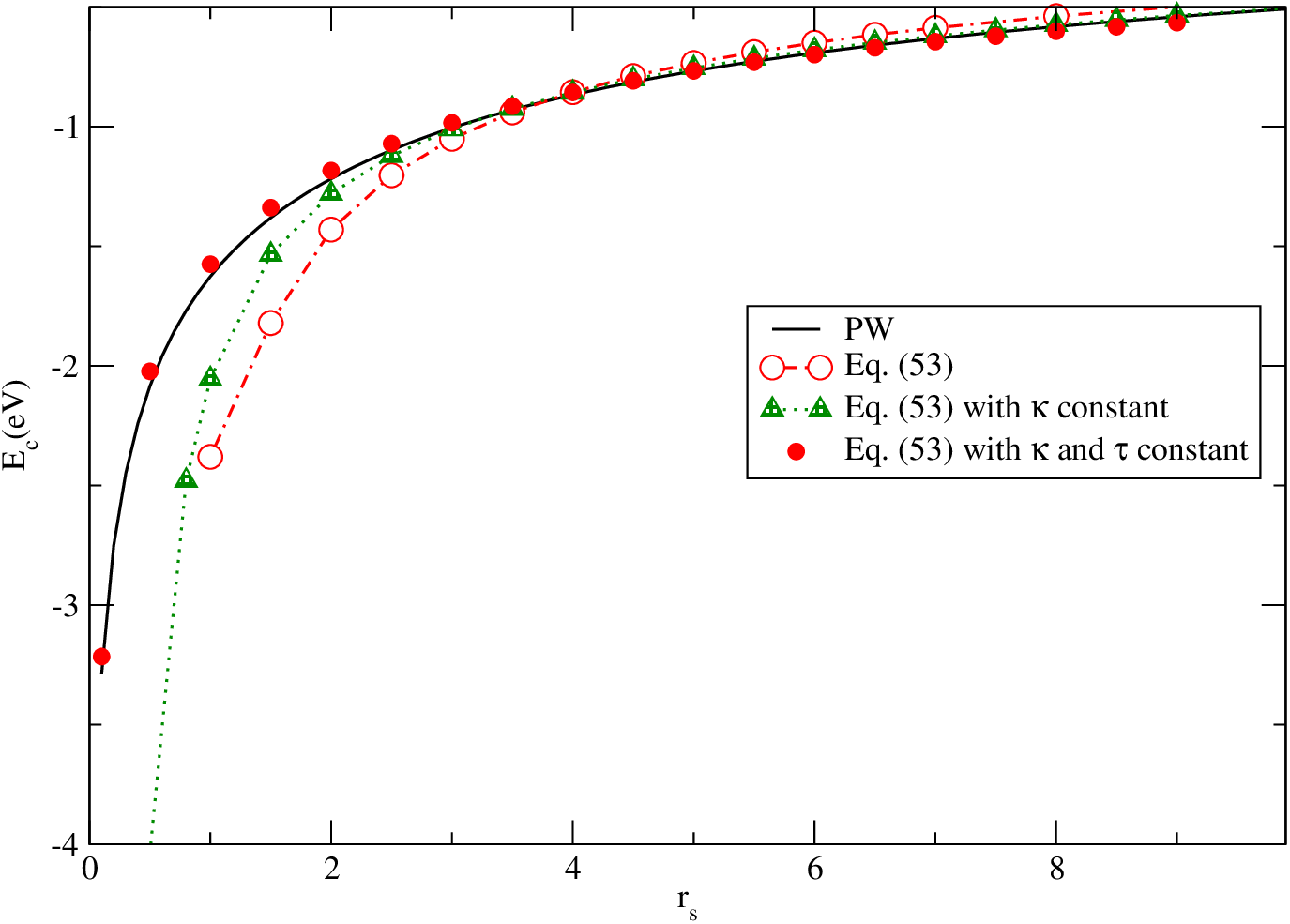}
\caption{Integrated electron correlation energy computed for the Gaussian cut effective
  interaction (Eq.(\ref{62}). The cases studied are the one with core condition (\ref{70}) 
  (i.e.~Eqs.~(\ref{71}) and (\ref{72}) are fulfilled), then $\kappa$ is kept constant 
  (i.e.only Eq.~(\ref{72}) is solved), and finally both $\kappa$ and $\tau$ are kept constant.
  The constant values of $\kappa$ and $\tau$ are determined
  solving for the core conditions at $r_s=4$. The reference
  values, fully drawn curve, are those of the Perdew-Wang parameterization \cite{PW}.} 
\label{fxc}
\centering
\end{figure}

\begin{figure}[h!]
\centering
\includegraphics*[width=18cm, angle=0,clip]{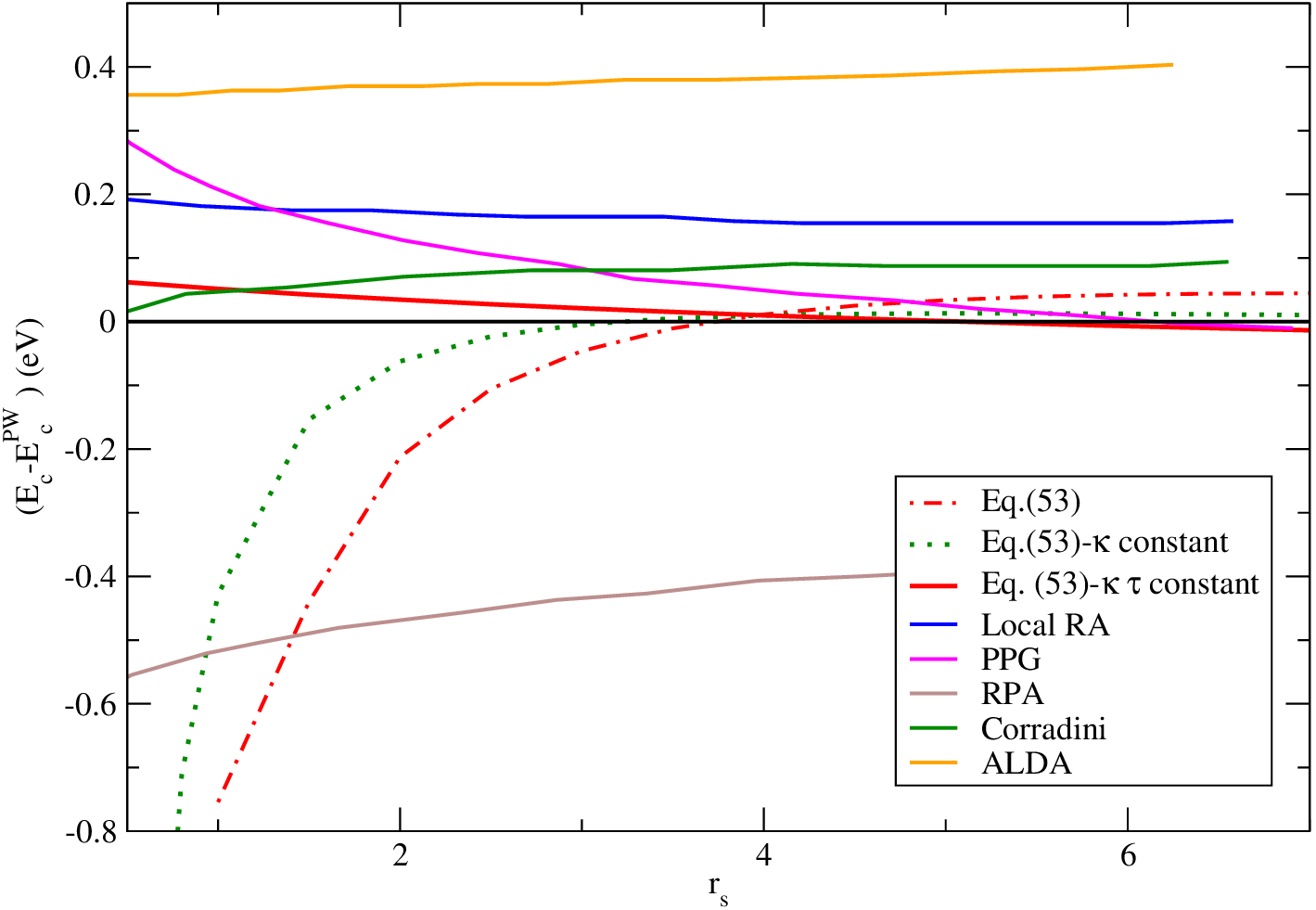}
\caption{Deviation of the  electron correlation energy computed using the
  approximated core conditions in this work and using other
  approximations from the literature (see the text for definitions)
  with respect to the Perdew-Wang parameterization\cite{PW}.}  
\label{difxc}
\centering
\end{figure}

\end{document}